# Efficient generation of optical bottle beams

Yuzhe Xiao[1], Zhaoning Yu[1,2], Raymond A. Wambold[1], Hongyan Mei[1], Garrett Hickman[2], Randall H. Goldsmith[3], Mark Saffman[2], and Mikhail A. Kats[1,2,4*]

[1]Department of Electrical and Computer Engineering, University of Wisconsin-Madison, Madison, Wisconsin 53706, USA

[2]Department of Physics, University of Wisconsin-Madison, Madison, Wisconsin 53706, USA

[3]Department of Chemistry, University of Wisconsin-Madison, Madison, Wisconsin 53706, USA

[4]Department of Materials Science and Engineering, University of Wisconsin-Madison, Madison, Wisconsin 53706, USA

*Correspondence to: mkats@wisc.edu.

**Abstract:** Optical bottle beams can be used to trap atoms and small low-index particles. We introduce a figure of merit for optical bottle beams, specifically in the context of optical traps, and use it to compare optical bottle-beam traps obtained by three different methods. Using this figure of merit and an optimization algorithm, we identified optical bottle-beam traps based on a Gaussian beam illuminating a metasurface that are superior in terms of power efficiency than existing approaches. We numerically demonstrate a silicon metasurface for creating an optical bottle-beam trap.

**Introduction**

Optical bottle beams are characterized by localized light-intensity minima at their foci that are surrounded by regions of higher light intensity in all directions, and thus can be used for trapping atoms and low-index particles [1]–[5]. Several approaches have been used to generate bottle-beam traps. One approach uses the interference of two optical beams with different beam parameters such that destructive interference results in a dark region at the mutual focus of the two beams [6]–[8]. This approach requires precise alignment of the two beams as well as control of their beam waists, amplitudes, and relative phases. Another approach uses a single optical beam with a functional optical component, such as a diffractive optical element [9], hologram [2], or spatial light modulator [10], [11]. Recently, metasurfaces have also been considered for bottle-beam generation [12]–[14]. Due to the compact size of metasurfaces, this approach has potential for chip-scale optical trapping.

However, it is difficult to compare these different trapping strategies. Here, we introduce a figure of merit (FoM) to quantify the efficiency of two different types of optical bottle-beam traps: a point trap (mostly for atoms) and a volume trap (mostly for small particles). Note that by "point", we mean that our design minimizes the optical field at a particular point instead of over some volume; nevertheless, any far-field optical bottle will always extend over some volume due to diffraction.

We then use parameter-sweep optimization to compare the best traps obtained via three different methods: (i) using destructive interference of two Gaussian beams combined on a beam splitter, (ii) using a single Gaussian beam incident on a metasurface whose transmission phase comes from the interference of two Gaussian beams, or (iii) using a single Gaussian beam incident on a metasurface whose transmission phase comes from an accelerating beam. Utilizing the freedom of design enabled by metasurface engineering, we identified bottle-beam traps with optimum performance in terms of power efficiency. As a proof of concept, we numerically demonstrate a finite-volume bottle-beam trap using a silicon metasurface.

**Evaluation of a bottle-beam trap**

First, we introduce two figures of merit (FoMs) to evaluate the quality and power efficiency of optical bottle-beam traps. As an example, the intensity profile of a bottle-beam trap in the $x$-$z$ plane is schematically shown in Fig. 1(a). This optical bottle-beam trap is rotationally invariant with respect to the $z$ axis and has a minimum intensity at the origin ($x = y = 0$). A good trap should satisfy the following three main criteria.

First, the light intensity surrounding the trap should be high. The trapping power or the trap depth can be quantified by the escape intensity $I_{escape}$, which we define as the smallest value among the maximum light intensity values along all of the straight lines emanating from the center of the trap. For the bottle-beam trap shown in Fig. 1(a), the max intensity is different along different directions due to the asymmetry of the trap: e.g., the max intensity is higher along the $x$ direction [$I_{1,max}$ in Fig. 1(b)] and lower along the $z$ direction [$I_{3,max}$ in Fig. 1(b)]. Therefore, $I_{escape} = I_{3,max}$ for this trap. Note that for particles that are not very small compared to the size of the trap, this definition may be insufficient.

Second, the light intensity inside the trap should be low since it is the contrast in the light intensity inside and outside the trap that determines the trapping power. In addition, for atom traps, a lower intensity inside the trap is favorable because it means less photon scattering, less laser-noise-induced heating, and therefore longer coherence times [15]. Note that we consider two cases: one trap chiefly for small particles, which has a certain volume where the mean intensity is minimized, and one trap chiefly for atoms, where the size of the trap may not matter as long as the trap has a central point with minimum intensity.

Lastly, the trap should be power efficient: the input power required for creating a given contrast of intensity inside and outside of the trap should be minimized. This condition is important for cases when the input laser power is the limiting factor (e.g., a single trap that needs very large trapping power or an array of traps created by a single laser [16]).

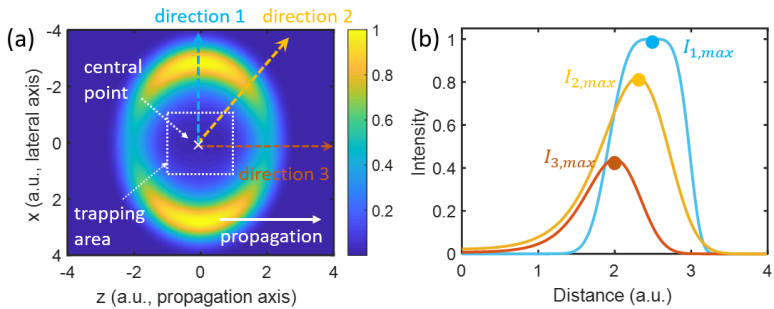

Fig. 1. (a) Schematic intensity profile of an optical bottle-beam trap in the $x$-$z$ plane. The beam propagates along the $z$ direction and is rotationally invariant along the $z$ axis. Intensity profiles along three different directions are shown in (b), where the escape intensity for the trap in (a) is $I_{escape} = I_{3,max}$.

Therefore, based on the above considerations, we define our FoM for optical bottle-beam traps as follows:

$$\eta = \frac{I_{escape} - I_0}{P} \cdot 10^{-9} \text{ m}^2, \tag{1}$$

Where $P$ is the power of the incident beam that creates the trap. For a small-particle "volume" trap, $I_0$ is the averaged intensity inside the trapping volume, while for a "point" trap for atoms, $I_0$ is the intensity at the central point of the trap. We incorporated an additional factor of $10^{-9}$ m$^2$ in our definition to make $\eta$ unitless, and such that $\eta$ has a magnitude on the order of 100 for traps of size of a few microns.

**Approaches and parameter-sweep optimization**

Many methods can be used to generate bottle-beam traps [2], [7], [9], [12]. In this work, we focus on three methods based on two different physical approaches. The first approach uses the destructive interference of two Gaussian beams [7], where the beams have the same focus location and similar intensity at the focus, but different waists and a $\pi$ phase difference at the focus, as illustrated in Fig. 2(a). Note the resulting bottle beam is rotationally invariant with respect to the $z$ axis and we only show the cross-section in the $x$-$z$ plane. Near the focus at $z = z_0$, there is a low-intensity trapping region due to the destructive interference of the two Gaussian beams. The radial extent of this trapping volume (along the $x$ axis) is roughly equal to the size of the smaller Gaussian beam and its extent along the propagation direction (the $z$ axis) is determined by the difference between the Gouy phase of these two beams due to their different waists [17].

To create such a Gaussian interference bottle-beam trap, one can use two Gaussian beams and a beam splitter using the setup schematically shown in Fig. 2(c) [7], which we denote as Method 1. Alternatively, one could use a single Gaussian beam incident on a metasurface designed to convert the Gaussian beam into a bottle beam, as shown in Fig. 2(d), which we denote as Method 2. In this case, the metasurface has a transmission-phase profile that matches the phase of two super-imposed Gaussians such as those in Method 1. Note that the metasurface we design here generates an optical bottle at a wavelength of 770 nm, 1 mm from the metasurface, with applications to an integrated single-photon source based on rubidium (Rb) atoms [18]. In the following calculations, we use Fresnel diffraction to simulate the fields of bottle-beam traps and only consider field profiles that are rotationally invariant. Therefore, the field propagation is performed using the following equation [19]:

$$E(r,z) = \frac{e^{-ikz}}{i\lambda z} 2\pi e^{-ik\frac{r^2}{2z}} \int_0^\infty \rho E(\rho, z=0) e^{-ik\frac{\rho^2}{2z}} J_0(-2\pi \rho r) d\rho \quad (2)$$

Here $J_0$ is the zero-order Bessel function of the first kind and is defined as:

$$J_0(x) = \frac{1}{2\pi} \int_0^{2\pi} e^{ix\cos(\theta)} d\theta \quad (3)$$

The electric field of two destructively interfering Gaussian beams with different waists ($w_{1,2}$) is

$$E(r,z) = G(1, w_1, r, z) - G(A, w_2, r, z) \quad (4)$$

where $G(A_0, w_0, r, z)$ is the electric field of a Gaussian beam with amplitude $A_0$ and waist $w_0$ [20]:

$$G(A_0, w_0, r, z) = A_0 \frac{w_0}{w(z)} \exp\left(-\frac{r^2}{w(z)^2}\right) \exp\left[-i\left(kz + k\frac{r^2}{2R(z)} - \psi(z)\right)\right] \quad (5)$$

Here, $k = 2\pi/\lambda$ is the wave vector, $w(z) = w_0\sqrt{1 + (z/z_R)^2}$ is the spot size where $z_R = \pi w_0^2/\lambda$ is the Rayleigh range, $R(z) = z[1 + (z_R/z)^2]$ is the radius of curvature, and $\psi(z) = \arctan(z/z_R)$ is the Gouy phase. Note that because our FoM is normalized to the input power, the absolute intensity does not matter. Therefore, we assume the first Gaussian has an amplitude of one and add a relative amplitude $A$ to the second Gaussian beam. For Method 1, the waist of two beams $w_1$, $w_2$, and relative amplitude $A$ fully determine the bottle beam. For Method 2, the phase transmission of the metasurface is determined by $w_1$, $w_2$, and $A$, while the trap also depends on the width of the incident Gaussian beam $R_0$.

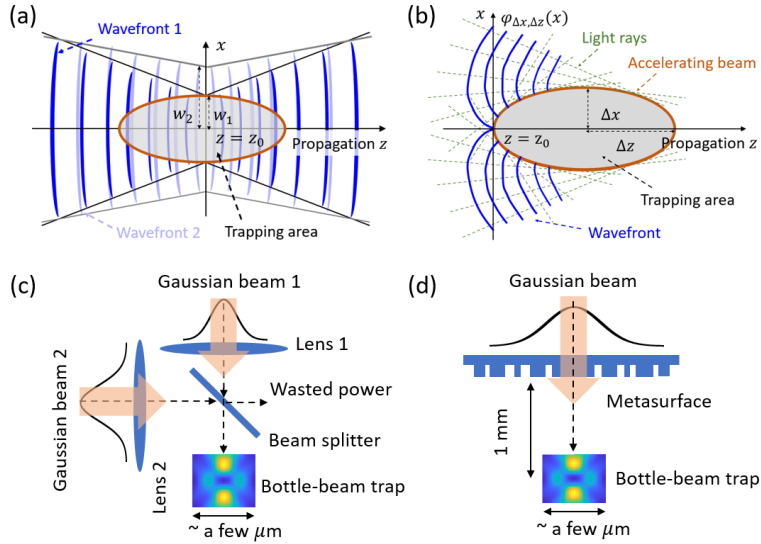

Fig. 2. (a) An optical bottle beam formed by the destructive interference of two Gaussian beams with different waists and a $\pi$ phase shift at the mutual focus. The radial extent (e.g., along the $x$ axis) of the low-intensity region (trapping volume) is roughly equal to the waist of the smaller beam, and the extent along the propagation direction ($z$ axis) is determined by the difference in the waist-dependent Gouy phase. (b) Design of an optical bottle beam by reverse phase retrieval of an accelerating beam. Starting from the bent

trajectory of an accelerating beam, one can find the rays that can be thought of as forming the beam, from which the wavefront can be constructed. Therefore, one can form a bottle beam by creating the corresponding phase response. (c) Schematic of a setup for creating a bottle beam using destructive interference of two Gaussian beams super-imposed using a beam splitter. (d) Schematic of a setup for creating a bottle beam using a single Gaussian beam incident on a metasurface with an appropriate phase response.

The second approach, i.e. Method 3, is based on reverse phase retrieval of an optical beam with a curved trajectory, a.k.a. an accelerating beam [21]. The basic idea is shown in Fig. 2(b). One starts with the trajectory of an accelerating optical beam (the top part (i.e., $x > 0$) of the curved trajectory in Fig. 2(b)), which defines the boundary of the trap. Then, one can form a bottle-beam trap by circularly wrapping the accelerating beam with respect to the $z$ axis. The resulting beam is rotationally invariant with respect to the $z$ axis. In the example shown here, the beam trajectory comprises the top half of an ellipse defined by the values of the two axes $\Delta x$ and $\Delta z$. According to the asymptotic theory of accelerating beams [12], light rays forming this beam are tangents of its trajectory. Using this geometrical relationship, one can find the wavefronts, which are perpendicular to the light rays (note that this can only trivially be done for $z > z_0$, because for $z < z_0$ there are two intersecting rays for every point). Then, the phase profile of this accelerating beam at the plane $z = z_0$, i.e., $\varphi_{\Delta x, \Delta z}(x, z = z_0)$, can be determined from the wavefronts [12]. We can then use $\varphi_{\Delta x, \Delta z}(x, z = z_0)$ to determine the required metasurface phase at $z = z_1$ to generate this trap, i.e., $\varphi_{\Delta x, \Delta z}(x, z = z_1)$. To back-propagate the field from $z_0$ to $z_1$ we assume a Gaussian field-intensity, where the Gaussian has width $r_0$ at $z = z_0$. We denote this method as Method 3.

We performed parameter-sweep optimizations to find the bottle-beam traps with the highest $\eta$ based on these three methods. The parameters that are relevant to the bottle-beam traps for these three methods are summarized in Table 1. As an example, we selected a free-space wavelength of 770 nm, which is used for blue-detuned trapping of Rb atoms [18], [22]. We optimized for two different traps: a volume trap of size 4 μm × 2 μm (in the $x$-$z$ plane) and a point trap. For the point trap, we observed that $\eta$ of the best trap increases for an optical system with a higher numerical aperture (NA). Therefore, we set the NA to be the same (0.7) for all three methods: an entrance pupil of radius 1 mm is placed 1 mm away from the trap. We swept all possible combinations of the parameters that generate bottle-beam traps via three methods. The best bottle-beam traps found using this parameter-sweep optimization are shown in Figs. 3(a-f), with the corresponding parameters listed in Table 1. The corresponding field profiles 1 mm away from the trap are shown in Sec. 1 in Supplementary Information.

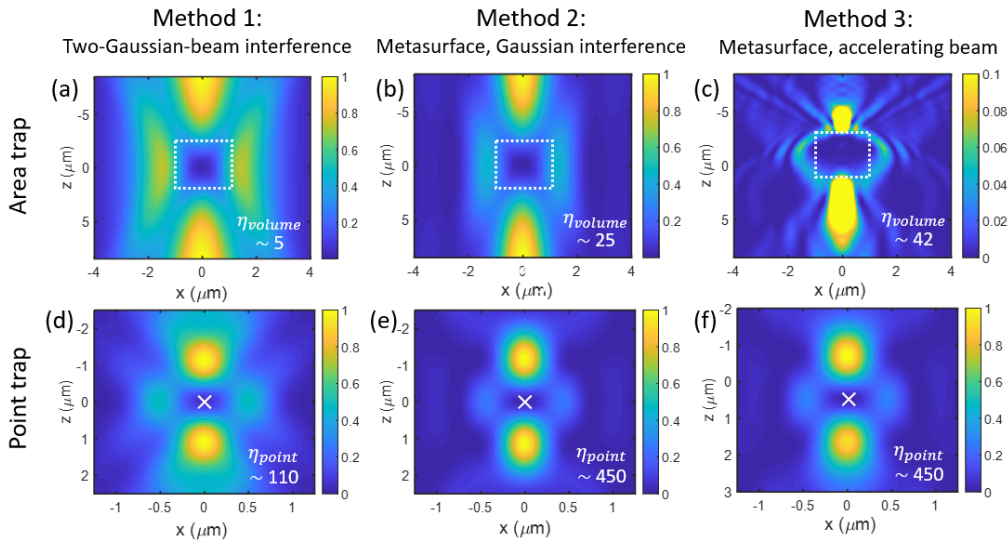

Fig. 3. Results of parameter-sweep optimization for generating optical bottle-beam traps at $\lambda_0 = 770$ nm based on three different methods: using the destructive interference of two Gaussian beams (Method 1),

illuminating a single Gaussian beam on an optical metasurface whose phase profile is used to mimic the two-Gaussian-interference beam (Method 2), and reverse phase retrieval from an accelerating beam (Method 3). Panels (a-c) depict the best volume traps targeting a trapping area of 4 µm × 2 µm (in the $x$-$z$ plane) for three different methods. Note that the intensity scale in (c) is different to better visualize the trap. (d-f) show the best point trap for a numerical aperture of 0.7 for three different methods.

|  | $A$ | $w_1$ ($\mu m$) | $w_2$ ($\mu m$) | $R_0$ ($\mu m$) | $\eta_{point}$ | $\eta_{volume}$ |
|---|---|---|---|---|---|---|
| Method 1, volume | 1.14 | 0.96 | 2.84 |  |  | 5 |
| Method 1, point | 0.73 | 0.23 | 0.76 |  | 110 |  |
| Method 2, volume | 0.61 | 0.69 | 1.32 | 225 |  | 25 |
| Method 2, point | 0.57 | 0.19 | 0.35 | 1100 | 450 |  |
|  | $\Delta x$ ($\mu m$) | $\Delta z$ ($\mu m$) | $r_0$ ($\mu m$) | $R_0$ ($\mu m$) | $\eta_{point}$ | $\eta_{volume}$ |
| Method 3, volume | 1.8 | 2.59 | 2.60 | 825 |  | 42 |
| Method 3, point | 0.24 | 0.65 | 0.73 | 860 | 450 |  |

Table 1. Values of the parameters that were used to generate the bottle-beam traps shown in Figs. 3(a-f). All traps are generated at $\lambda_0 = 770$ nm. The "volume" traps are for an area of 4 µm × 2 µm (in the $x$-$z$ plane) and the "point" traps are generated for a numerical aperture of 0.7.

For the optimized bottle-beam traps for microparticles (i.e., a "volume" trap) in Figs. 3(a-c), Method 1 (using two-Gaussian interference) [Fig. 3(a)] has an $\eta$ much smaller than Method 2 and 3 (using a metasurface with a phase profile generated from Gaussian interference and accelerating beams, respectively) [Figs. 3(b, c)] though the trap profiles in Figs. 3(a) and (b) look similar. This is mainly due to the destructive interference of the two incident Gaussian beams used in Method 1, which is not a power-efficient way to synthesize the bottle beam. In addition to the much larger $\eta$, the metasurface-based methods enable integration of the optical system into a more compact form factor because one does not need to precisely control and align two beams. Among the metasurface-based traps, Method 3 [Figs. 3(c)] creates a trap with a larger $\eta$ than Method 2 [Fig. 3(b)] due to the following features: a cleaner and darker trapping volume, a narrower barrier, and a more uniform trapping intensity along different directions. These features lead to a higher $\eta$ of the bottle-beam trap in Fig. 3(c).

For the optimized bottle-beam trap for atoms (i.e., a "point" trap) for an optical system with NA of 0.7 in Figs. 3(d-f), Method 1 [Fig. 3(d)] again has a much smaller $\eta$ than Method 2 and 3 [Figs. 3(e, f)] due to the inefficient use of incident power. But all three optimized point traps [Figs. 3(d-f)] look very similar, especially the two metasurface-based traps [Figs. 3(e, f)] which look almost identical. The fact that the optimized traps generated by metasurface with two different designs are the same indicates that this point bottle-beam trap might be close to the global optimum of traps that can be generated by a Gaussian beam and a metasurface. Indeed, as will be discussed later, we further run global optimizations and the best point trap found is quite close to the ones shown in Figs. 3(e, f).

**Point-by-point gradient-ascent optimization**

One advantage of the metasurface method is that one has the freedom of creating an arbitrary phase or amplitude response. Therefore, when designing metasurfaces for generating bottle-beam traps, one does not need to follow the phase profile from either the two-Gaussian-beam interference or the reverse phase retrieval of an accelerating beam approach, which may not necessarily generate the best bottle-beam traps. Therefore, we built a point-by-point gradient ascent optimizer to see whether the bottle-beam traps generated using metasurface in Fig. 3 can be improved further.

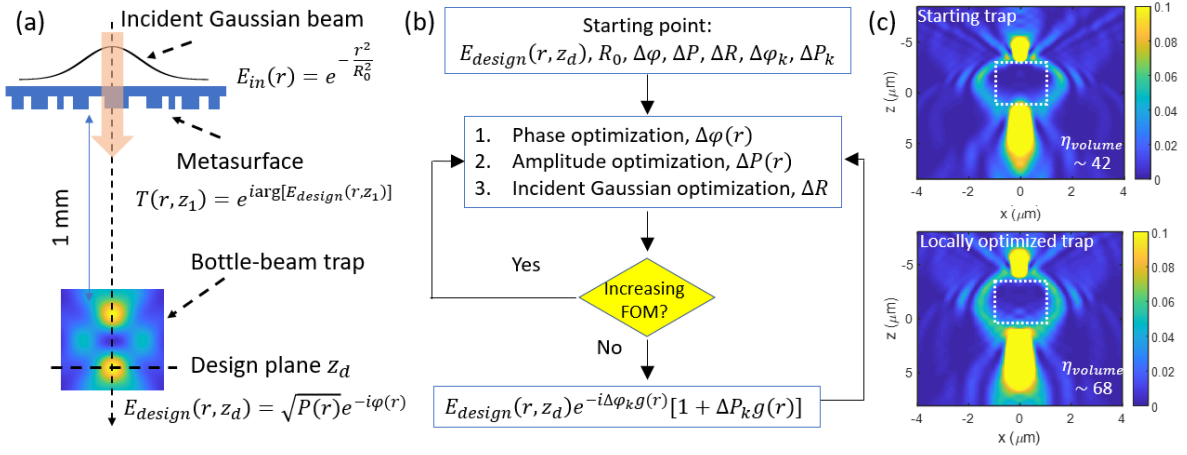

Fig. 4. (a) Schematic of the gradient-ascent optimization setup. First, the target field at the plane $z = z_d$ near the trap center is back-propagated to the metasurface at $z = z_1$. where the phase of $E_{design}(r, z_1)$ is used as the transmission phase of the metasurface. The metasurface is illuminated with a collimated Gaussian beam and the transmitted field produces the bottle-beam trap. (b) Flowchart of the point-by-point gradient-ascent optimization. Each iteration goes through three steps: optimization of the phase and amplitude of $E_{design}(r, z_d)$, as well as the incident Gaussian beam width. If the $\eta$ stops increasing, the program attempts to jump out of the local optimum by adding random variations to both the amplitude and phase of $E_{design}(r, z_d)$. (c) Profiles of the input [top, the trap in Fig. 3(c)] and locally optimized (bottom) volume bottle-beam traps, where the $\eta$ of the bottle-beam trap is increased by more than 60%.

The working distance of the metasurface is 1 mm in our design, which leads to a mm-scale size of the metasurface due to the diffraction of the beam over the 1-mm distance from the trap to the metasurface plane. Directly optimizing each pixel of such a mm-scale metasurface involves a very heavy computational load. Therefore, we choose a plane [$z = z_d$ in Fig. 4(a)] which is closer to the trap to optimize the electric field $E_{design}(r, z_d)$. Note that we only consider structures that are rotationally invariant with respect to the $z$ axis. We first back-propagate the starting field $E_{design}(r, z_d)$ to the metasurface located at $z = z_1$ using Eq. 2. The phase profile of $E_{design}(r, z_1)$ is used as the transmission phase of the metasurface, whose amplitude transmission is assumed to be one everywhere. Then the metasurface is illuminated with a collimated Gaussian beam such that the transmitted electric field is $e^{-r^2/R_0^2} e^{i \arg[E_{design}(r, z_1)]}$, where $R_0$ is the size of the incident Gaussian beam. The transmitted field is then propagated forward using Eq. 2 to generate the bottle-beam trap.

The optimization program modifies the phase and amplitude of $E_{design}(r, z_d)$, as well as the width of the illumination Gaussian beam to find better traps. At each iteration, the optimization consists of three steps [Fig. 4(b)]. In the first step, the phase of each pixel $j$ of $E_{design}(r_j, z_d)$ is changed by $\Delta\varphi$, and the corresponding change in $\eta$ of the trap $\Delta\eta_j$ is calculated. Then, the design field is updated as $E_{design}(r, z_d) = E_{design}(r, z_d) e^{-i\Delta\varphi f(r)}$, where $f(r_j) = \Delta\eta_j / \max(|\Delta\eta|)$. The function $f(r)$ indicates the direction of steepest ascent [23]. In the second step, the amplitude of each pixel $j$ of $E_{design}(r_j, z_d)$ is changed by $\Delta P$, and the corresponding change in $\eta$ of the trap $\Delta\eta_j$ is calculated. Then the design field is updated as $E_{design}(r, z_d) = E_{design}(r, z_d)[1 + \Delta P f(r)]$. Finally, in the third step, the width of the incident Gaussian beam is varied by [$-2\Delta R$, $-\Delta R$, $\Delta R$, $2\Delta R$]. Then the width that returns the largest $\eta$ is chosen for the next iteration.

The optimization iteration continues as long as $\eta$ keeps increasing. If $\eta$ stops increasing, the program attempts to escape the local optimum by adding random perturbations to both the amplitude and phase as $E_{design}(r, z_d) e^{-i\Delta\varphi_k g(r)}[1 + \Delta P_k g(r)]$, where $\Delta\varphi_k$ and $\Delta P_k$ are the perturbation phase and amplitude, and $g(r)$ is a smoothed random function of $r$.

We performed such optimization starting with some of the bottle-beam traps generated via parameter-sweep optimization [Figs. 3(b, c, e, f)]. In these optimizations, the value of $\Delta\varphi_k$ and $\Delta P_k$ were set to be small, such that this should be viewed as a local optimization around the parameter-sweep-optimized bottle-beam traps in Fig. 3. For each of the trap shown in Figs. 3(b, c, e, f), we ran 500 simulations with a randomly selected $\Delta\varphi$ and $\Delta P$. Within each of these simulations, the program runs 1000 iterations of the optimization algorithm as shown in Fig. 4(b). The trap with the highest $\eta$ was then chosen as the output of each simulation.

We found that for the bottle-beam traps shown in Figs. 3(b, e, f), the improvement in $\eta$ using the local gradient-ascent optimization is within 5% (Sec. 2 in Supplementary Information), meaning that the initial bottle-beam traps found by parameter-sweep optimization are quite close to local optimum. However, for the bottle-beam trap in Fig. 3(c), local optimization was able to find a better trap with an improvement of about 60% in $\eta$. The starting and the optimized trap profiles are shown in Fig. 4(c). The corresponding field profiles at the metasurface are shown in Sec. 2 in Supplementary Information.

The optimization process illustrated in Fig. 4(b) searches for a better trap locally by starting from a particular design field and following the gradient of steepest ascent in $\eta$, escaping local optima by re-starting from a field that is lightly perturbed from the same design field. This program can also be adjusted to search for good traps globally by escaping the local optimum by re-starting from a random field (completely unrelated to the parameter-sweep-optimized designs) every time $\eta$ stops increasing. We performed such global searches both for the volume and point bottle-beam traps, and for each type we ran 1000 simulations with randomly selected $\Delta\varphi$, $\Delta P$, $\Delta\varphi_k$, and $\Delta P_k$. In each of these 1000 simulations, the program ran 1000 iterations of the optimization algorithm as shown in Fig. 4(b). The best bottle-beam traps found from these global searches with the corresponding intensity and phase profile of the electric field at the metasurface are shown in Fig. 5.

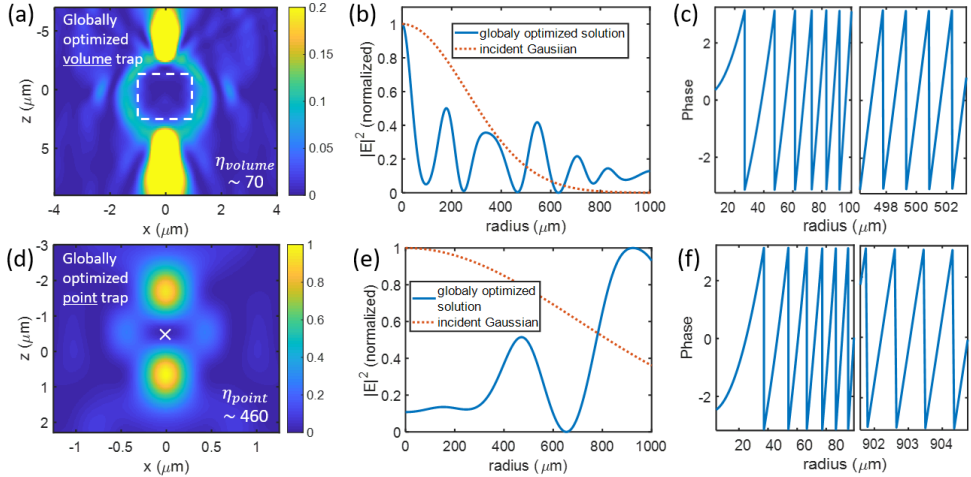

Fig. 5. The best (a) volume and (d) point bottle-beam traps found by "global" gradient-ascent optimization [NA = 0.7 in (d)]. The corresponding intensity and the phase profiles of (a) and (d) at the metasurface plane are shown in (b, c) and (e, f).

For a volume bottle-beam trap with a size of 4 μm × 2 μm (in the $x$-$z$ plane), the best trap found by global gradient-ascent optimization has an $\eta$ of 70 [Fig. 5(a)], which is quite close to the result from local gradient-ascent optimization shown in the bottom part of Fig. 4(c). In addition, the intensity distributions of these two bottle-beam traps are also similar. For a point bottle-beam trap, the best trap found by global optimization has an $\eta$ of 460 [Fig. 5(d)], which is quite close to the value found by parameter-sweep optimization in Figs. 3(e, f). In fact, these three traps look almost identical. This observation further indicates that this trap is probably close to the global optimum.

**Demonstration with a silicon metasurface**

Here we design an optical metasurface that generates a volume bottle-beam trap, targeting a size of 4 μm × 2 μm (in the $x$-$z$ plane) at the wavelength of 770 nm. There has recently been substantial progress in the development of dielectric metasurfaces [24]–[26], including metasurfaces based on silicon for applications in the visible [27], [28] and near-infrared spectral regions [29]. Here, we consider crystalline silicon nano cylinders on top of a fused-silica substrate as the basic metasurface elements. At 770 nm, the loss of crystalline silicon is relatively small [30] and fused silica is transparent, so the metasurface can have transmission close to one. After running simulations on a single unit cell for combinations of different silicon cylinder heights and unit-cell periods, we set the height of the silicon cylinder to be 360 nm and the period to be 330 nm. In Fig. 6(a), we plotted the transmitted field amplitude and phase as a function of the diameter of the silicon cylinder. A full $2\pi$ phase difference can be obtained while maintaining a roughly constant amplitude transmission by varying the diameter from 115 to 220 nm. Note here, the transmitted field amplitude is larger than 1 V/m because the source is a plane wave with amplitude of 1 V/m launched within the fused-silica substrate.

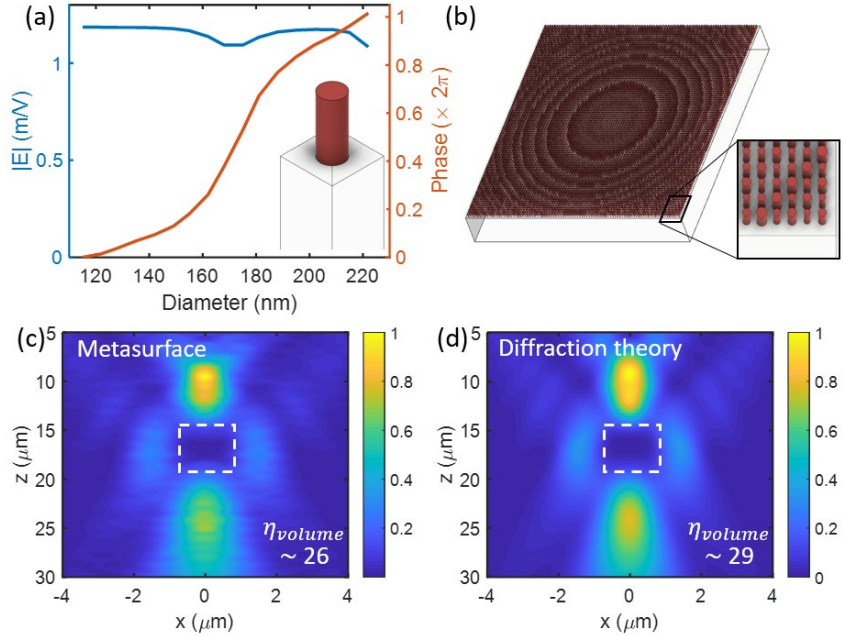

Fig. 6. Demonstration of a volume bottle-beam trap using a crystalline silicon metasurface. (a) Transmitted field amplitude and phase as a function of the silicon cylinder diameter. The height of the cylinder is 360 nm, with a period of 330 nm, sitting on top of a fused-silica substrate. The source is a plane wave at a free-space wavelength of 770 nm launched from inside the silica substrate. (b) Simulated silicon metasurface. The dimension of the metasurface is 30 μm × 30 μm and the design is radially symmetric. The inset is a portion of the metasurface near its edge. (c) Simulated bottle-beam trap profile using the metasurface in (b) via full 3D FDTD simulation. The metasurface was located at $z = 0$ $\mu m$ and was illuminated with a Gaussian beam of width of 5 $\mu m$. $\eta$ of the generated bottle-beam trap is 26. (d) Calculated bottle-beam trap using diffraction theory. The incident field at $z = 0$ $\mu m$ is the product of the transmission phase of the metasurface in (b) and a Gaussian beam of 5 $\mu m$ wide. This bottle-beam trap has an $\eta$ of 29.

The metasurface optimized above has a size on the order of 1 mm$^2$, for which it is very hard to run full 3D finite-difference time domain (FDTD) simulations. As a proof of concept, we simulated a much smaller metasurface of size 30 μm × 30 μm, as shown in Fig. 6(b). The phase profile of this metasurface was obtained by reverse phase retrieval of an accelerating beam such that a bottle-beam trap with a size of 4 μm × 2 μm (in the $x$-$z$ plane) is formed at about 15 $\mu m$ away from the metasurface when illuminated by a Gaussian beam 5 $\mu m$ wide. The simulation was performed using Lumerical FDTD, and the calculated bottle-beam trap profile is shown in Fig. 6(c), with $\eta \sim 26$. As a comparison, we also performed the corresponding calculation using diffraction theory via Eq. 2. Note that the bottle-beam traps generated via the metasurface approach shown earlier in this work are also calculated using diffraction theory. In the diffraction-theory calculation shown in Fig. 6(d), only the phase profile of the metasurface is used, with the transmission assumed to be one everywhere. The simulated trap using diffraction theory is plotted in Fig. 6(d) and has $\eta \sim 29$. The trap profiles in Figs. 6(c) and (d) are quite similar, with the difference comes from the fact that the metasurface used in Fig. 6(c) has a slightly non-uniform transmission.

**Conclusion**

We introduced figures of merit (FoMs) to evaluate the performance of optical bottle-beam traps for atoms and small low-index particles, in terms of trapping effectiveness for a given incident laser power. We compared the best bottle-beam traps from three different methods: the first method uses the destructive interference of two Gaussian beams combined on a beam splitter, the second and third methods use a single incident Gaussian beam transmitted through a metasurface with a deliberately designed phase profile. We found that the use of a metasurface (or other efficient wavefront-converting device) significantly reduces the incident power needed to create a bottle-beam trap compared to the method that combines two beams on a beam splitter. The use of metasurfaces enables arbitrary control of the transmitted phase as a function of position, and these degrees of freedom can be used to further improve bottle-beam traps. By optimizing the metasurface transmission phase, we identified designs with probable global optima in trap performance. We also performed full-wave simulations of a metasurface based on silicon pillars that generated a bottle-beam trap, illustrating how optimal bottle-beam designs can be implemented in practice.

**Acknowledgements:** This material is based upon work supported by the National Science Foundation under Grant No. PHY-1839176 and Grant No. 2016136. RW acknowledges support through the DoD SMART program and a scholarship from the Directed Energy Professional Society. GH and MS acknowledge support from U.S. Army Research Laboratory Center for Distributed Quantum Information through Cooperative Agreement No. W911NF-15-2-0061.

# Supplementary Information for
# Efficient generation of optical bottle beams

Yuzhe Xiao[1], Zhaoning Yu[1,2], Raymond A Wambold[1], Hongyan Mei[1], Garrett Hickman[2], Randall H Goldsmith[3], Mark Saffman[2], and Mikhail A. Kats[1,2,4*]

[1]Department of Electrical and Computer Engineering, University of Wisconsin-Madison, Madison, Wisconsin 53706, USA
[2]Department of Physics, University of Wisconsin-Madison, Madison, Wisconsin 53706, USA
[3]Department of Chemistry, University of Wisconsin-Madison, Madison, Wisconsin 53706, USA
[3]Department of Materials Science and Engineering, University of Wisconsin-Madison, Madison, Wisconsin 53706, USA
*Correspondence to: mkats@wisc.edu.

## 1. Field profiles from parameter-sweep optimization

Here we show the field profiles at 1 mm away from the trap that generate the traps in Fig. 3. More specifically, the far-field profiles that generate the traps in Figs. 3 (a-c) and Figs. 3 (d-f) are shown in Fig. S1 and S2, respectively.

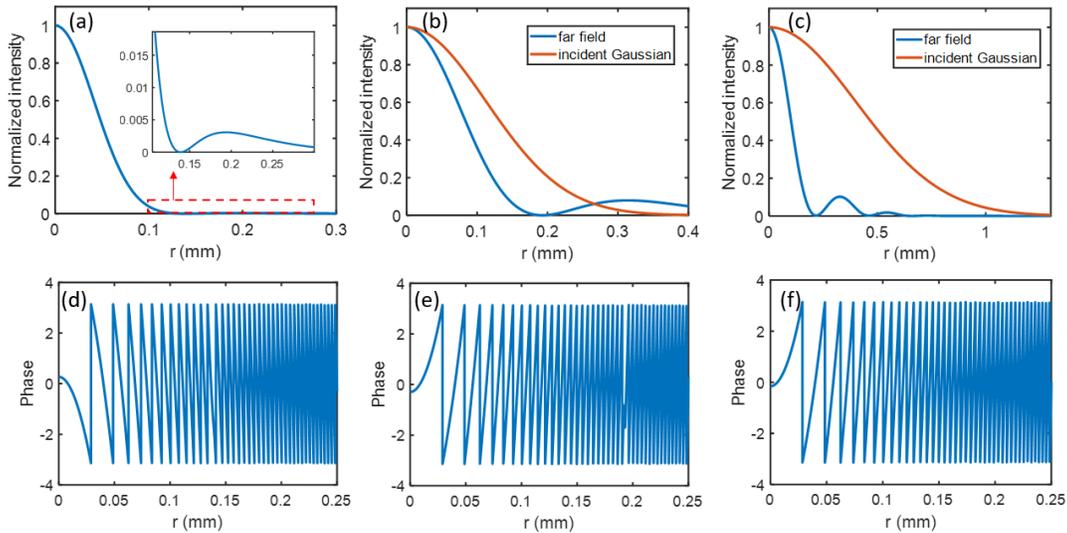

Fig. S1. (a-c): Intensity profiles of the far field as well as the incident Gaussian that generate the area bottle trap shown in Figs. 3 (a-c). (d-f): Corresponding phase profiles of the far field.

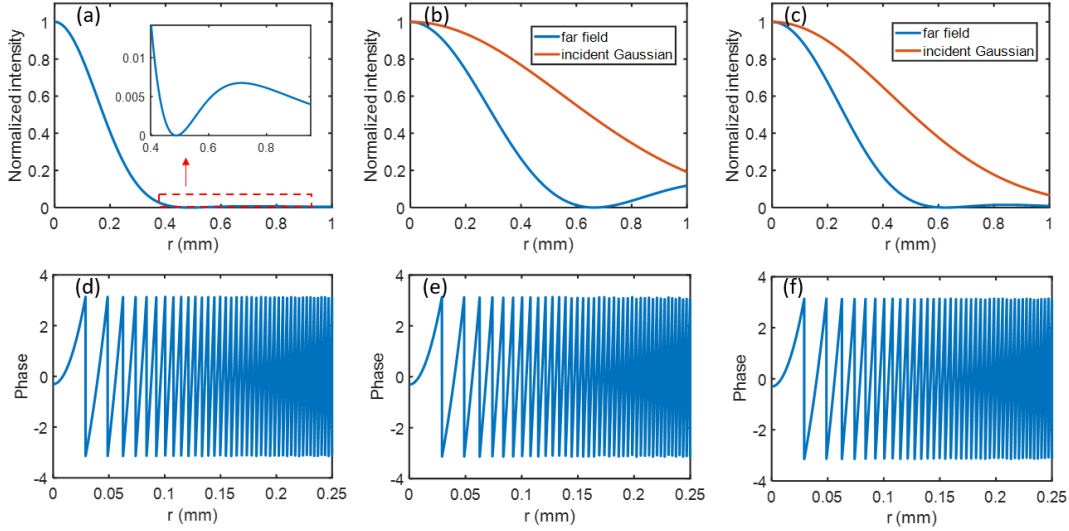

Fig. S2. (a-c): Intensity profiles of the far field as well as the incident Gaussian that generate the point bottle trap shown in Figs. 3(d-f). (d-f): Corresponding phase profiles of the far field.

## 2. Local optimization of the bottle traps

We found that the traps shown in Figs. 3 (b, e, f) are quite close to a local optimum because the best traps found from local optimization are quite close to these traps. As an illustration, Fig. S3 shows the result of one of the 500 simulations for the trap in Fig. 3(b). As shown in Fig. S3(a), the highest FOM is almost the same as the starting point (the red dotted line). The best trap from this local optimization shown in Fig. S3(b) looks almost identical to the one shown in Fig. 3(b). The field profiles as well as the incident Gaussian of the optimized trap are very close to the starting point.

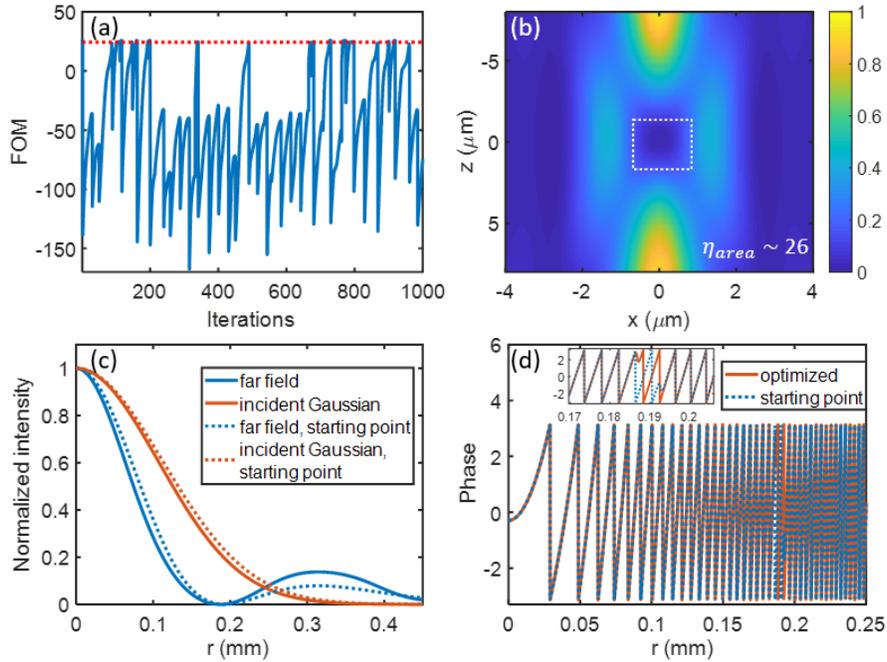

Fig. S3. (a): Evolution of FOM for different iterations of the local optimization of one simulation for the trap shown in Fig. 3(b). The FOM of the starting trap is indicated by the red dotted line. Out of 500 different simulations, the highest FOM is 26, which is very close to the trap in Fig. 3(b). (b): The trap profile of the best trap found in this simulation, which is almost identical to the trap shown in Fig. 3(b). (c) and (d) compare the far field profiles as well

as the incident Gaussian for the best trap found by parameter-sweep optimization in Fig. 3(b) (dotted lines) and local optimization (solid lines).

For the trap shown in Fig. 3(c), the local optimization was able to find a trap with a FOM much higher. As an illustration, Fig. S4 shows the result of one of the 500 simulations for the trap in Fig. 3(c). As shown in Fig. S3(a), the highest FOM shows an improvement of about 60% over the starting point (the red dotted line). The best trap from this local optimization shown in Fig. 4(c) looks similar to the starting trap. The field profiles as well as the incident Gaussian of the optimized trap are close to the starting point.

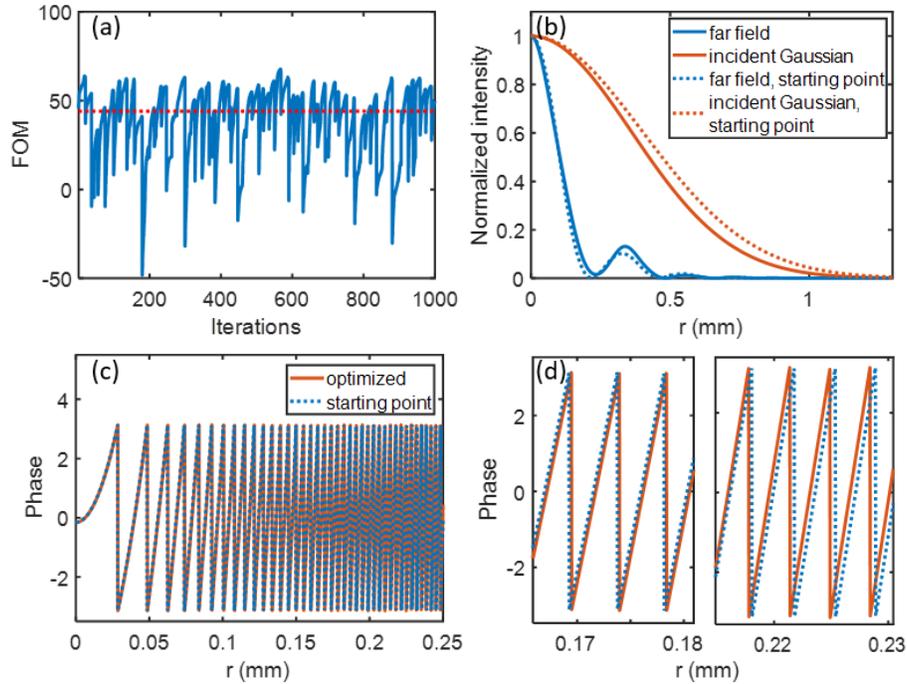

Fig. S4. (a): Evolution of FOM for different iterations of the local optimization of one simulation for the trap shown in Fig. 3(c). Out of 500 different simulations, the highest FOM is about 68, which shows an improvement of 60% from the starting value. (b-d) compare the far field profiles as well as the incident Gaussian for the best trap found by parameter-sweep optimization in Fig. 3(c) (dotted lines) and local optimization (solid lines).

## 3. Global optimization of the bottle traps

Here we compare the best area bottle beam traps found from the global optimization with those found from the local optimization.

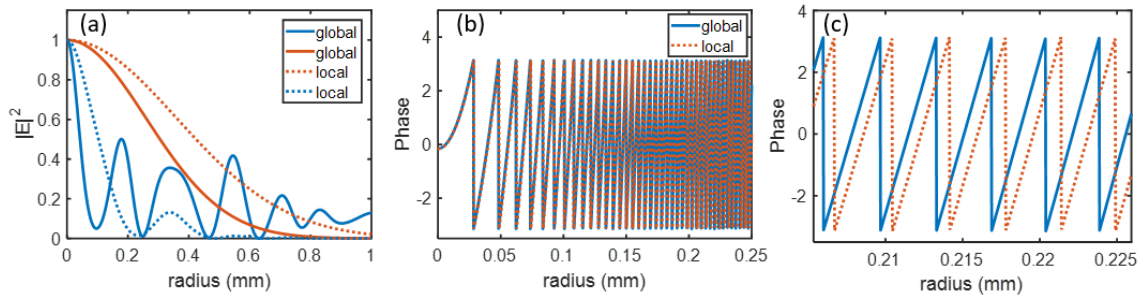

Fig. S5. (a): Comparison of the far field intensity profile as well as the incident Gaussian for the area bottle trap found from local (dotted lines) and global optimization (solid lines). Phase profiles for these two traps are plotted in (b), with the zoomed in portion near 0.22 mm shown in (c).